\documentstyle[pra,  aps, float]{revtex}
\begin{document}

\draft
\date{\today}

\title{Bethe logarithms for the $1^1\mbox{S}$, $2^1\mbox{S}$ and
$2^3\mbox{S}$ states of helium and helium-like ions} 
\author{Jonathan
D. Baker$^1$ Robert C. Forrey$^2$ Malgorzata Jeziorska$^3$ John
D. Morgan III$^4$}


\address{
$^1$National Institute of Standards and Technology, Gaithersburg, MD 20899 \\
$^2$Penn State University, Berks-Lehigh Valley College,
	Reading, PA 19610-6009\\
$^3$Department of Chemistry, University of Warsaw, Pasteura 1, 
        02-093 Warsaw, Poland \\
$^4$Department of Physics and Astronomy, University of Delaware, 
	Newark, DE 19716}
\maketitle
\begin{abstract}
We have computed the Bethe logarithms for the $1^1\mbox{S}$, $2^1\mbox{S}$ 
and $2^3\mbox{S}$ 
states of the helium atom to about seven figure-accuracy using a 
generalization of a method first developed by Charles Schwartz.   
We have also calculated the Bethe logarithms for the helium-like ions Li$^{+}$, Be$^{++}$,
O$^{6+}$ and S$^{14+}$ for all three states 
to study the $1/Z$ behavior of the results.  The Bethe logarithm
of H$^-$ was also calculated with somewhat less accuracy.  The use of 
our Bethe logarithms for the excited states of neutral helium, instead of 
those from Goldman and Drake's first-order $1/Z$-expansion, reduces by 
several orders of magnitude the discrepancies between the theoretically 
calculated and experimentally measured ionization potentials of these states.
\end{abstract}


    Ever since the invention of quantum mechanics, the helium atom has 
served as an important testing-ground for our understanding of fundamental
physics.  In 1929 Hylleraas' calculation of the binding energy of the 
non-relativistic helium atom Hamiltonian showed that Schroedinger's 
formulation of quantum mechanics provided a quantitatively accurate 
description of not just two-body but three-body systems \cite{EAH:1929}.  
During the 1950's,
with the advent of fast digital computers, calculations by Kinoshita 
\cite{TK:1957} and Pekeris \cite{CLP:1958} of not only the non-relativistic 
binding energy but also of 
relativistic corrections of O($\alpha^2$) Rydberg greatly improved 
the agreement between theory and experiment, and showed that the estimation 
of O($\alpha^3$) Rydberg effects arising from quantum electrodynamics 
was important for obtaining agreement between theory and experiment at 
the level of 1 part in 10$^6$
or better.  During the 1960's and 1970's the variational techniques employed by 
Pekeris on the lowest states of singlet and triplet symmetry were extended 
to a wide range of excited states of the helium atom 
\cite{FP:1966} \cite{APS:1971}.  
During the 1980's, 
with the advent of two-photon spectroscopy with counterpropagating laser 
beams, which can be used to 
eliminate the 1st-order Doppler shift due to the thermal motion of the
atoms, it became possible to measure 
the wavelengths for transitions between excited states of the helium 
atom with a precision of 1 part in 10$^9$ or better \cite{GB:1982}.  
Though numerous 
examples of excellent agreement between theory and experiment in a wide 
variety of contexts leave no reasonable doubt that quantum electrodynamics 
is the correct theory for describing the interactions of charged particles
at low energies, the extraordinary accuracy recently achieved in 
high-precision measurements on the helium atom poses a challenge to 
theorists to develop computational techniques capable of matching such 
accuracies.   Since $\alpha^3$ is of order 10$^{-6}$, it is clear that 
the coefficient of the lowest-order QED corrections needs to be evaluated 
with a relative accuracy of 10$^{-3}$ or better, and the effects of 
contributions with higher powers of $\alpha$ must also be estimated, 
to match the experimental accuracy of 1 part in 10$^9$ or better.

For a helium atom or helium-like ion of atomic number $Z$, 
the leading O($\alpha^3$) Rydberg contribution to the Lamb shift is given
by the expression \cite{PKK:1957}

\begin{eqnarray*} 
E_{L,2}=\frac{8}{3}Z \alpha^3 {\em \Psi}_0^2(0) 
\left[ 2 \ln{\left(\frac{1}{\alpha}\right)}-
\ln{\left(\frac{k_0}{\mbox{Ry}}\right)} +\frac{19}{30}
\right]\mbox{Ry},
\end{eqnarray*}
where the so-called Bethe logarithm \cite{HB:1947} is defined by an 
infinite and slowly-convergent sum over all bound and continuum eigenstates:
\begin{eqnarray*}
\lefteqn{\ln (k_{0}/\mbox{Ry}) = \frac{\beta}{\cal D} = } \\ & &
 \frac{ \sum_{n}\mid \mbox{$\left\langle {\em \Psi}_n \left|
 \right. \right.$}{\bbox p} \mbox{$\left. \left. \right| {\em \Psi}_0
 \right\rangle$} \mid^{2}(E_{n}-E_{0}) \ln |E_{n}-E_{0}|}
 {\sum_{n}\mid \mbox{$\left\langle 
    {\em \Psi}_n \left| \right. \right.$}{\bbox p}
 \mbox{$\left. \left. \right| {\em \Psi}_0 \right\rangle$}
 \mid^{2}(E_{n}-E_{0})},
\end{eqnarray*}
Here ${\bbox p}$ is the sum of single-particle momentum operators 
(${\bbox p}\!=\!\sum_i{\bbox p}_i$)
and ${\em \Psi}_0$ is an eigenfunction with eigenvalue $E_0$ of 
the Hamiltonian $H$ of the atom.  For simplicity, we 
assume that $H$ is the nonrelativistic Hamiltonian of an atom with 
atomic number $Z$, with a point nucleus of infinite mass:
 
$$
H = T + V = \sum_{i} p_i^2/2 - Z \sum_{i} 1/r_i + \sum_{i>j} 1/r_{ij} , 
$$
which has the important and useful property that it is unitarily equivalent
to the scaled Hamiltonian 
$$
Z^2 \left( \sum_{i} p_i^2/2 
          - \sum_{i} 1/r_i + (1/Z) \sum_{i>j} 1/r_{ij} \right) ,
$$
which after division by $Z^2$ tends to a well-defined limit as 
$Z \rightarrow \infty$.
(The effects of the reduced mass $\mu = m_e M_N/(m_e + M_N)$ due to the 
finiteness of the 
nuclear mass $M_N$ are subsequently included by scaling by appropriate powers 
of $\mu / m_e$, and the negligible effect of the `mass-polarisation' term  
$M_N^{-1} \sum_{i>j} {\bf p}_i \cdot {\bf p}_j$ on 
the Bethe logarithm is here ignored.)  
With the help of the closure relation 
$\sum_{n} | n \rangle (E_n - E_0) \langle n | \, = \, H - E_0$,
the commutation relation 
$(H - E_0) {\bbox p} {\em \Psi}_0 \, = i (\nabla V) {\em \Psi}_0$, 
an integration by parts, and Gauss' Law 
($\nabla^2 V = 4 \pi Z \sum_i \delta^{(3)}({\bf r}_i)$), 
the denominator ${\cal D}$ is easily evaluated:
\begin{eqnarray*} 
{\cal D} = \mbox{$\left\langle {\em \Psi}_0 \left| \right. \right.$}
{\bbox p} \cdot (H-E_0) {\bbox p} \mbox{$\left. \left. \right| {\em \Psi}_0
\right\rangle$} = 2 \pi Z {\em \Psi}_0^2(0),
\end{eqnarray*}
but the logarithmic factor makes the numerator $\beta$ much harder
to evaluate.  Even for a very simple one-electron system such as the hydrogen
atom, $\beta$ cannot be evaluated in closed form, though several
rapidly convergent methods can be used to evaluate it to high
accuracy \cite{SPG:1984}, \cite{BHM:1989}, \cite{GWFD:1990}, \cite{RCF:1993},
and the Bethe logarithm of the electronic ground state of H$_2^+$ was recently 
evaluated numerically \cite{BJMK:1992}.  For a two-electron system such 
as the helium atom, whose unknown wavefunction $\Psi_0$ must be represented
by an expansion in a large basis set, the numerical challenges are even
more daunting.

In the early 1960's C. Schwartz recast the numerator as
integral over the virtual photon energy $k$ \cite{CS:1961},
\begin{eqnarray} 
\lefteqn{\beta = \lim_{K \rightarrow \infty} \left( \rule{0ex}{2.5ex}
\right. \mbox{$-K\left\langle {\em \Psi}_0 \left|
\right. \right.$}{\bbox p} \cdot {\bbox p}
\mbox{$\left. \left. \right| {\em \Psi}_0 \right\rangle$} +{\cal D}
\ln(K) + \nonumber } \\
 & & \int_0^K k \; dk 
 \mbox{$\left\langle {\em \Psi}_0 \left| \right. \right.$} 
  {\bbox p} \cdot (H-E_0+k)^{-1} {\bbox p}
\mbox{$\left. \left. \right| {\em \Psi}_0 \right\rangle$}
\left. \rule{0ex}{2.5ex} \right) ,
\label{eq:beta}
\end{eqnarray}
and thereby replaced the insuperable difficulties associated with
accurately summing over an infinite number of
bound and continuum eigenstates of $H$ with the more tractable difficulty of
numerically integrating an accurate representation of the matrix element of the
resolvent $(H-E_0+k)^{-1}$ for small, intermediate and large values of
$k$.  When $k$ is very large, Schwartz found it sufficient to
approximate the matrix element with a simple asymptotic formula. For
smaller values of $k$, the action of the resolvent is solved
explicitly as the solution of a system of linear equations in a
suitable basis with $p$-wave symmetry. For intermediate $k$ the convergence 
was greatly improved by including a single function which has
the same leading-order asymptotic behavior as the
true solution as  $k \rightarrow \infty$.


Despite growing problems with
the numerical linear dependence of his basis as the number of basis
functions was increased, Schwartz was able to compute for the 1$^1$S 
ground state of the neutral helium atom a
Bethe logarithm of 4.370(4) Rydbergs, which yielded a
theoretical ionization potential for this state in
agreement with the best experimental values available at that time, 
and which remained unsurpassed until very recently.

The results presented in this letter were generated by an approach
very similar to that used by Schwartz, in which the integral in
Eq. (\ref{eq:beta}) is split into a low $k$ region
$\beta_{\mbox{\scriptsize L}}$ and a high $k$ region
$\beta_{\mbox{\scriptsize H}}$.  The counterterms in
Eq. (\ref{eq:beta}) are then brought inside the integral  to
cancel explicitly the divergent behavior at large $k$:
\begin{eqnarray} 
 \beta & = & \beta_{\mbox{\scriptsize H}} + \beta_{\mbox{\scriptsize L}} 
 = \int_1^\infty \frac{dk} {k} \mbox{$\left\langle {\em \Psi}_0
\left| \right. \right.$} {\bbox p} \cdot (H-E_0)
\mbox{$\left. \left. \right| \psi_{\mbox{\scriptsize H}}(k)
\right\rangle$} \nonumber \\ && + \int_0^1 k \; dk \left(
\mbox{$\left\langle {\em \Psi}_0 \left| \right. \right.$} {\bbox p}
\mbox{$\left. \left. \right| \psi_{\mbox{\scriptsize L}}(k)
\right\rangle$}- \frac{\mbox{$\left\langle {\em \Psi}_0 \left|
\right. \right.$}{\bbox p} \cdot {\bbox p}
\mbox{$\left. \left. \right| {\em \Psi}_0 \right\rangle$}}{k} \right),
\label{eq:hillint} 
\end{eqnarray}
where $\psi_{\mbox{\scriptsize L}}(k)$ and $\psi_{\mbox{\scriptsize H}}(k)$
are solutions of the equations
\begin{eqnarray} 
&& \label{eq:psil} (H-E_0+k) \; \psi_{\mbox{\scriptsize L}}(k) =
{\bbox p} {\em \Psi}_0, \\ && (H-E_0+k) \; \psi_{\mbox{\scriptsize H}}(k) 
= (H-E_0){\bbox p} {\em \Psi}_0.
\label{eq:psih}
\end{eqnarray}
Since $H$ possesses overall rotational symmetry, the solutions 
$\psi_{\mbox{\scriptsize L}}(k)$ and $\psi_{\mbox{\scriptsize H}}(k)$
have a total angular momentum quantum number which can differ by only 
$\pm1$ from that of ${\em \Psi}_0$.  In this work ${\em \Psi}_0$ has 
$S$-symmetry, so 
$\psi_{\mbox{\scriptsize L}}(k)$ and $\psi_{\mbox{\scriptsize H}}(k)$
have $P$-symmetry.  

An elegant derivation of Eq. (\ref{eq:hillint}) can be found in the
work of Forrey and Hill \cite{RCF:1993}, which examines Schwartz's method 
from a fresh perspective and provides many useful computational techniques.  
We evaluate the two integrals in
Eq. (\ref{eq:hillint}) numerically, using the procedure described by
Forrey and Hill, computing the matrix element of the resolvent at each
integration knot by solving variationally for $ \psi_{\mbox{\scriptsize L}} $ or
$\psi_{\mbox{\scriptsize H}}$ in Eq. (\ref{eq:psil}) and
Eq. (\ref{eq:psih}). 
When $k$ is very large, we use the asymptotic approximation \cite{CS:1961}
\begin{eqnarray} 
\lefteqn{\mbox{$\left\langle {\em \Psi}_0 \left| \right. \right.$}
{\bbox p} \; (H-E_0) \mbox{$\left. \left. \right|
\psi_{\mbox{\scriptsize H}}(k) \right\rangle$} = } \nonumber \\ & &
\frac{2 Z {\cal D}}{k} \left[\sqrt{2k}-Z \ln(k)+C+\frac{D}{\sqrt{k}} +
\cdots \right].
\label{eq:asymp}
\end{eqnarray}
The constants $C$ and $D$ have been computed in closed form only for
the hydrogen atom; in this work they are estimated by extrapolating 
the values generated by the solution of 
Eq. (\ref{eq:psih}) at successive integration knots.  This equation
was solved explicitly at each successive knot, running in the
direction of increasing $k$, until the relative difference between
successive extrapolated estimates of $C$ was roughly 1\%. For larger $k$
the resulting asymptotic formula was used. For the helium ground state
our estimates of $C$ and $D$ are 4.988(1) and -18.8(3) respectively,
with the errors resulting mainly from extrapolation uncertainty.
These estimates can be compared with the value 5.18 computed by Schwartz
\cite{CS:1961} for $C$ and the value -20$\pm$3 he assumed for
$D$.

The non-relativistic wavefunction ${\em \Psi}_0$ was computed variationally 
using our modification \cite{BHM:1989}, \cite{MORGAN:1984} of the  
basis set first developed by Frankowski and Pekeris \cite{FP:1966}, which
exploits knowledge of the analytic structure of the true wavefunction
at the 2- and 3-particle coalescences to improve the convergence of the 
variational trial function to the exact unknown wavefunction:

\begin{eqnarray*} 
 & & {\em \Psi}_0 = \sum_{\nu} c_{\nu} (\phi_{\nu}( s, t, u) \pm 
  \phi_{\nu}( s, - t, u)) \\ 
 & & \phi_{\nu}(s,t,u) = s^n t^l u^m (\ln s)^{j} e^{-as+ct}
\end{eqnarray*}
where $s$, $t$, and $u$ are the Hylleraas coordinates defined by 
$s\!=\!r_1\!+\!r_2$, $t\!=\!r_2\!-\!r_1$ and $u\!=\!r_{12}$ and 
the $\pm$ sign is chosen so that the product of ${\em \Psi}_0$
and the spin function is antisymmetric under exchange of the electrons.

Our bases for representing
$\psi_{\mbox{\scriptsize L}}(k)$ and $\psi_{\mbox{\scriptsize H}}(k)$  
include functions of four different types.  The $k$-independent functions 

$$\chi^{(1)}_{\nu}={\bbox r}_1\phi_{\nu}(s,t,u) \pm {\bbox r}_2
\phi_{\nu}(s,-t,u)$$
together with the single function $\chi^{(2)} = {\bbox p}{\em \Psi}_0$
provide a good solution space for small $k$.  

For large $k$ the solution $\psi_{\mbox{\scriptsize H}}(k)$ becomes 
concentrated in $k$-dependent regions of configuration space for which 
one electron is very close to the nucleus and the other electron is much
further away, so it is essential to use explicitly $k$-dependent basis 
functions. Of primary importance is the `Schwartz function' $\chi^{(3)}$, an 
approximate solution of
Eq. (\ref{eq:psih}) that reproduces the first two terms in the
asymptotic expansion in Eq. (\ref{eq:asymp}):
\begin{eqnarray*} 
\chi^{(3)}=\left({\bbox p}_1 \frac{\exp{(-\sqrt{2k}\,r_1)}-1}{r_1}{\em
\Psi}_0 \right) \pm \left({\bbox r}_1 \leftrightarrow {\bbox r}_2
\right).
\end{eqnarray*}
To help approximate that part of $\psi_{\mbox{\scriptsize H}}(k)$ which is 
orthogonal to the `Schwartz function',
we also use a fourth set of functions $\chi^{(4)}_{\nu}$,   
which are symmetrized sums 
of products of single-variable Laguerre functions 
${\cal L}_i(R_j)\!=\!L_i(R_j)e^{-R_j/2}$
of the three perimetric coordinates 
$R_1\!=\!r_1\!+\!r_2\!-\!r_{12}$, $R_2\!=\!r_1\!-\!r_2\!+\!r_{12}$, 
and $R_3\!=\!-r_1\!+\!r_2\!+\!r_{12}$:
\begin{eqnarray*}
  \chi^{(4)}_{\nu}=({\bf r}_1\, {\cal L}_p(aR_1) \, {\cal L}_q(bR_2) \, 
{\cal L}_r(cR_3) ) \pm ({\bbox r}_1 \leftrightarrow {\bbox r}_2). 
\end{eqnarray*}
Combinations of the exponential parameters $a$, $b$, and $c$ can be chosen 
to reflect the strong `in-out' correlation in 
$\psi_{\mbox{\scriptsize H}}(k)$ for large $k$.  
For any $k$ the overlap matrix elements for these basis functions are very 
small or zero far from the main diagonal, which enables us to avoid the severe 
problems with numerical linear dependence which prevented Schwartz from using a 
large basis of functions of the form of powers of $r_1$, $r_2$, $r_{12}$ 
times a highly asymmetrical exponential of  $r_1$ and $r_2$.  
We set $a=b+c$ to eliminate from $\chi^{(4)}_{\nu}$ any exponential 
$r_{12}$-dependence, 
which would complicate the evaluation of  matrix elements between these 
functions and the other types of basis functions.
Analytic considerations \cite{RCF:1993} suggest that the
integrand is optimised if $b \approx (2k)^{1/4}$ and $c \approx Z$.
We coarsely search the
parameter space in the neighborhood of these values of $b$ and $c$
seeking to maximize the two integrands of Eq. (\ref{eq:hillint})
in accordance with the variational principle described in \cite{RCF:1993}. 

The calculation of $\beta_{\mbox{\scriptsize L}}$ was fast and
straightforward. In this case $k$ was small enough that there was
no need to include explicit $k$-dependence in the basis. 
$\chi^{(3)}$ was omitted altogether, and a single {\it
average} value of the parameter $b$ was used in the $\chi^{(4)}_{\nu}$
functions, independent of the value of $k$ at a particular integration
knot.  We solved for $\psi_L (k)$ in a basis with
92 $\chi^{(1)}_{\nu}$
functions, the $\chi^{(2)}$ function, and 120 $\chi^{(4)}_{\nu}$
functions.  The parameters $b$ and $c$ were varied to
maximize the integrand.  Changes in the integrand due to small
variations in $b$ and $c$ were used to assess convergence.
 
The $\beta_{\mbox{\scriptsize H}}$ integral was computationally
expensive, primarily because including the `Schwartz function' $\chi^{(3)}$
requires evaluating algorithmically complicated
matrix elements. Since $\chi^{(3)}$ is intended primarily to accelerate 
the convergence for very large $k$, and since over half of the knots
in our integration scheme correspond to $k<40$, we chose to omit
$\chi^{(3)}$ from the basis for knots below $k\approx40$.
At each node we solved for 
$\psi_{\mbox{\scriptsize H}}$ in a basis
consisting of 92 $\chi^{(1)}_{\nu}$ functions, the $\chi^{(2)}$ function,
the $\chi^{(3)}$ function (for high $k$), and 220 $\chi^{(4)}_{\nu}$
functions. We then recomputed the solution of Eq. (\ref{eq:psih})
after first reducing the number $N$ of $\chi^{(4)}_{\nu}$ functions
in the existing matrices to study convergence of the integrand. 
A simple polynomial fit
in the variable $1/N$ was applied to the sequence of results with 
$N=$ 220, 165, 120, and 84 to
generate the values of the Bethe logarithms in this letter. 
The error associated with the finiteness of the basis for 
$\psi_{\mbox{\scriptsize H}}$ was taken as the entire difference 
between the extrapolated value
and the value corresponding to $N=220$.

Other sources of numerical error arise from the numerical
integration itself (for which there are good analytic error bounds
\cite{RCF:1993}), and the finiteness of the basis 
used to approximate
${\em \Psi}_0$. The latter error is assumed to be
comparable to the relative error in ${\cal D}$ in all cases. For
neutral helium, independent runs with less accurate representations of
${\em \Psi}_0$ indicate that this estimate of this error is somewhat
conservative. The numerical integration was parametrized to keep the
absolute error in $\beta_{\mbox{\scriptsize H}}$ and
$\beta_{\mbox{\scriptsize L}}$ below $10^{-8}$.  The results of 
independent calculations carried out for neutral helium with a coarser mesh
were consistent with the analytic error bound.

The uncertainties assigned to the Bethe logarithms in this letter are
the sums of the uncertainty due to extrapolation of
$\psi_{\mbox{\scriptsize H}}$ and the uncertainty due to approximation
of ${\em \Psi}_0$ in a finite basis. The uncertainty in
$\psi_{\mbox{\scriptsize L}}$ and the numerical integration error
bounds are negligible by comparison.

Our Bethe logarithms for the $1^1\mbox{S}$, $2^1\mbox{S}$, and
$2^3\mbox{S}$ states are listed in Tables \ref{grdoz}, \ref{singoz},
and \ref{tripoz}, respectively.  The values of $k_0$ have been divided
by $Z^2$ to illustrate their approach to the hydrogenic limit
as $Z$ becomes large.  Scaled values of the nonrelativistic binding
energy $E_{\mbox{nr}}/Z^2$ and ${\cal D}/Z^4$ are also listed to
provide some measure of the accuracy of ${\em \Psi}_0$.  Uncertainties
in ${\cal D}$ were computed by comparison with highly accurate results
for $<\!\delta({\bf r}_1)\!>$ provided by Drake \cite{GWFD:pc}.

The exact hydrogenic limits of $\ln(k_0/\mbox{Ry})$\cite{GWFD:1990}, 
\cite{GWFD:1993}
and of  $E_{\mbox{nr}}$ and ${\cal D}$ are displayed in the bottom 
row of each table, labeled by $1/\infty$ (exact). Immediately above 
the bottom row, in the row labeled $1/\infty$, we list the
hydrogenic values and the corresponding uncertainties computed using the
method described in this letter with $1/Z = 0$ so that the $1/r_{12}$ 
term is removed from the Hamiltonian. 

For $Z=1$ the Hamiltonian $H$ has a single bound state of  $1^1\mbox{S}$
symmetry.  As $Z \rightarrow 1$ from above all the singly excited bound states 
of a two-electron ion disappear into the continuum as the `outer' electron
moves infinitely far away.  Hence as 
$Z \rightarrow 1$ from above, the energies and all other finite-range 
properties of the states should tend toward those for a single hydrogen 
atom in its ground state with $Z=1$.   The approach of the Bethe 
logarithms and other properties toward their hydrogenic values  
as $Z \rightarrow 1$ from above is visible in Tables II and III 
for the  $2^1\mbox{S}$ and $2^3\mbox{S}$ states, respectively. 

We have fit our ionic results to the $1/Z$ expansion developed by 
Goldman and Drake
\cite{GWFD:1983}
\begin{eqnarray} 
\nonumber & & \ln(k_0/\mbox{Ry}) = C_0 + C_1 \, /Z + C_2 \, /Z^2 +
 \cdots \\ & & C_0= \ln 2 + 2 \ln Z + \ln(k_H/\mbox{Ry}),
\end{eqnarray}
where $\ln(k_H/\mbox{Ry})$ is the weighted sum of the two hydrogenic
Bethe logarithms corresponding to the state. Table \ref{overZ}
displays the results of a three parameter polynomial fit for $C_1$,
$C_2$, and $C_3$ using data for $Z=4, \ 8,\ \mbox{and} \ 16$.  The
listed uncertainties come from the formal propagation of error through
the regression formula and do not include truncation errors from
higher-order terms in the expansion.


Our results for the $1^1\mbox{S}$ state and the $2^1\mbox{S}$ states
of neutral helium are in complete agreement with the recent
calculations of Korobov and Korobov\cite{VK:1998}. The most accurate
previous value of the Bethe logarithm of the $2^3\mbox{S}$ state 
came from Goldman and Drake's 1st-order $1/Z$
expansion \cite{GWFD:1993}, \cite{GWFD:1983}. A numerical comparison 
of results for neutral helium appears in Table \ref{compb}.

Preliminary values of our Bethe logarithms 
for the $1^1\mbox{S}$, $2^1\mbox{S}$ \cite {BABS:1993}
and $2^3\mbox{S}$ states of helium were used in a recent
comparison of theory and experiment by Drake and Martin
\cite{GWFD:1998}.  The values in this letter make slight corrections
to the theoretical ionization energies of the $2^1\mbox{S}$ and the
$2^3\mbox{S}$ levels in that work, while the $1^1\mbox{S}$ state is
unaffected. Modifying ``{\it Bethe log cor.}''  contribution in
Drake and Martin's Table II to include the values in this letter yields
the theoretical results in Table \ref{ion}, which are compared with 
results from several recent experiments \cite{KSEE:1996}, \cite{SDB:1998}, 
\cite{WL:1991}, \cite{CJS:1992}, \cite{CD:1997}.

We are indebted to P.J. Mohr for helpful discussions related 
to this work and for his assistance in securing resources at NIST.
All numerical results in this letter were generated in the
fall of 1998 on either the NIST J40 IBM RS/6000 SMP machine or on the
IBM SP2, also at NIST.\footnote{Certain commercial equipment,
instruments, or materials are identified in this paper to foster
understanding. Such identification does not imply recommendation or
endorsement by the National Institute of Standards and Technology, nor
does it imply that the materials or equipment identified are
necessarily the best available for the purpose.}
    We would also like to acknowledge R.N. Hill for contributing
several useful ideas for setting up and performing the numerical
integration over virtual photon energy $k$. We thank G.W.F. Drake for
helpful discussions at an earlier stage of this work, for kindly
providing us with unpublished data from his work on helium-like ions
\cite{GWFD:1988}, and also for performing additional calculations to
facilitate our estimation of the uncertainty in ${\cal D}$.  We also
thank W.C. Martin for helpful discussions and Janine Shertzer and Tony 
Scott for their assistance and advice with the evaluation of integrals.  
We are also grateful to V.I. Korobov for keeping us informed of his
calculation of the Bethe logarithms.
Some computer
runs with an earlier version of this program were performed on an
RS/6000 system at the University of Washington kindly made available
to us by W.P. Reinhardt, and also on the SP-2 system at the Cornell
Theory Center.  
This work was supported by NSF grants PHY-8608155 and PHY-9215442 and
by a NIST Precision Measurement Grant to J.D. Morgan at the University
of Delaware, and by an NRC Postdoctoral Fellowship held by J.D. Baker
at NIST.
J.D. Morgan thanks the Institute for
Theoretical Atomic and Molecular Physics at Harvard University, and
its previous director, A. Dalgarno, for support in 1989-90 and 1992.
J.D. Morgan and J.D. Baker thank D. Herschbach and the members of his
research group at Harvard University for their hospitality.
which has greatly facilitated this work.  
They also thank the Institute for
Nuclear Theory at the University of Washington for providing support
in the spring of 1993. J.D. Morgan is further indebted to C.J. Umrigar
and M.P. Teter of the Cornell Theory Center for sabbatical support in
1995. 

\newpage

\begin{table}
\caption{$\ln(k_0/(Z^2\mbox{Ry}))$, $E_{\mbox{nr}}/Z^2$ and ${\cal
D}/Z^4$ for the $1^1\mbox{S}$ state. }
\label{grdoz}
\begin{tabular}{llll} 
1/$Z$ & $\ln(k_0/(Z^2\mbox{Ry})\rule{0ex}{3ex})$ & $E_{\mbox{nr}}/Z^2$
(a.u.) & ${\cal D}/Z^4$ (a.u.) \\ \tableline  
1/1 & 2.992\,97(5) & -0.527\,751\,015\,308 & 2.067\,80(4)      \\ 
1/2 & 2.983\,864(2) & -0.725\,931\,094\,259 & 2.843\,815\,67(5) \\ 
1/3 & 2.982\,624(2) & -0.808\,879\,268\,074 & 3.189\,069\,9(1)  \\ 
1/4 & 2.982\,503(1) & -0.853\,472\,889\,901 & 3.376\,853\,2(1)  \\ 
1/8 & 2.982\,948(2) & -0.924\,321\,798\,793 & 3.677\,270\,31(4) \\ 
1/16 & 2.983\,448(1) & -0.961\,551\,275\,290 & 3.835\,858\,39(2) \\ 
1/$\infty$ & 2.984\,128\,6(7) & -1.0 & 4.0 \\ 
1/$\infty$ (exact) & 2.984\,128\,556 & -1.0 & 4.0 \\
\end{tabular}
\end{table}

\begin{table}
\caption{$\ln(k_0/(Z^2\mbox{Ry}))$ , $E_{\mbox{nr}}/Z^2$ and 
${\cal D}/Z^4$ for the $2^1\mbox{S}$ state. }
\label{singoz}
\begin{tabular}{llll} 
1/$Z$ & $\ln(k_0/(Z^2\mbox{Ry})\rule{0ex}{3ex})$ & $E_{\mbox{nr}}/Z^2$
 (a.u.) & ${\cal D}/Z^4$ (a.u.) \\ \tableline 
1/1 (exact limit) & 2.984\,128\,556 & -0.5               & 2.0    \\ 
1/2 & 2.980\,115(1) & -0.536\,493\,511\,514 & 2.056\,896\,21(2) \\ 
1/3 & 2.976\,362(2) & -0.560\,097\,416\,177 & 2.103\,163\,60(3) \\ 
1/4 & 2.973\,976(1) & -0.574\,054\,618\,459 & 2.132\,593\,60(7) \\ 
1/8 & 2.969\,797(4) & -0.597\,793\,082\,931 & 2.185\,583(1)     \\ 
1/16 & 2.967\,459(3) & -0.610\,956\,015\,708 & 2.216\,320\,3(6)  \\ 
1/$\infty$ & 2.964\,977\,7(4) & -0.625 & 2.25 \\  
1/$\infty$ (exact) & 2.964\,977\,593 & -0.625 & 2.25 \\
\end{tabular}
\end{table}

\begin{table}
\caption { $\ln(k_0/(Z^2\mbox{Ry}))$ , $E_{\mbox{nr}}/Z^2$ and ${\cal
D}/Z^4$ for the $2^3\mbox{S}$ state. }
\label{tripoz}
\begin{tabular}{llll} 
1/$Z$ & $\ln(k_0/(Z^2\mbox{Ry})\rule{0ex}{3ex})$ & $E_{\mbox{nr}}/Z^2$
(a.u.) & ${\cal D}/Z^4$ (a.u.) \\ \tableline   
1/1 (exact limit) & 2.984\,128\,556 & -0.5     & 2.0            \\  
1/2 & 2.977\,742(1) & -0.543\,807\,344\,559 & 2.074\,008\,93(2) \\  
1/3 & 2.973\,852(1) & -0.567\,858\,596\,952 & 2.124\,087\,184(9) \\ 
1/4 & 2.971\,735(1) & -0.581\,072\,911\,861 & 2.152\,566\,566(2) \\ 
1/8 & 2.968\,414(2) & -0.602\,260\,114\,376 & 2.199\,147\,9(4) \\   
1/16 & 2.966\,705(1) & -0.613\,440\,895\,056 & 2.224\,062\,6(2) \\  
1/$\infty$ & 2.964\,977\,6(2) &   -0.625                & 2.25            \\
1/$\infty$ (exact) & 2.964\,977\,593 &-0.625          &  2.25            \\
\end{tabular}
\end{table}

\begin{table}
\caption{Coefficients of the $1/Z$ expansion using a 3 parameter
fit. The exact values of $C_1$ are due to Drake \protect \cite{GWFD:1993}.}
\label{overZ}
\begin{tabular}{llll} 
$1/Z \rule{0ex}{3ex}$ coeff & $ 1^1\mbox{S} $ & $ 2^1\mbox{S} $ & $
2^3\mbox{S} $ \\ \tableline $C_1$ (exact) & -0.0123\,03(1)& $\
$0.040\,771(1) & $\ $0.027\,760(1) \\ $C_1$ & -0.0123\,2(5) & $\
$0.040\,78(10) & $\ $0.027\,73(5) \\ $C_2$ & $\ $0.0228(8) & -0.016(2)
& -0.001\,0(6) \\ $C_3$ & $\ $0.002(2) & -0.011(6) & -0.007(2) \\
\end{tabular}
\end{table}

\begin{table}
\caption{Comparison of $\ln(k_0/\mbox{Ry})$ for neutral helium. The
uncertainty in the 1st-order $1/Z$ expansion due to uncalculated higher-order
terms could not be readily estimated until more exact calculations were done.}
\label{compb}
\begin{tabular}{lllll} 
State $\rule{0ex}{3ex}$ & 
           1st-order $1/Z$ expansion & Schwartz & Korobov & This work \\
\tableline 
$1^1\mbox{S}$ & 4.364(?)  & 4.370(4) & 4.370\,157\,9(5) & 4.370\,159(2) \\ 
$2^1\mbox{S}$ & 4.372(?)  & ---------  & 4.366\,409\,1(5) & 4.366\,409(1) \\ 
$2^3\mbox{S}$ & 4.365(?)  & ---------  & ------------------ & 4.364\,036(1)\\
\end{tabular}
\end{table}

\begin{table}
\caption{Ionization potentials, in MHz, as described in \protect
\cite{GWFD:1998}, but with theoretical values corrected slightly by
the results in this letter.  a: from \protect \cite{KSEE:1996}, b:
from \protect \cite{SDB:1998}, c: an average of values from \protect
\cite{WL:1991} and \protect \cite{CJS:1992}, d: from \protect
\cite{CD:1997}. }
\label{ion}
\begin{tabular}{llll} 
State $\rule{0ex}{3ex}$ & This work & Experiment & Difference \\
\tableline 
$1^1\mbox{S}$ & 5\,945\,204\,226(91) &
5\,945\,204\,238(45)$^a$ & $\ $12(102) \\ 
& & 5\,945\,204\,356(48)$^b$ & 130(103) \\ 
$2^1\mbox{S}$ & $\ \ $960\,332\,040.9(25.0) & $\ \
$960\,332\,041.01(15)$^c$ & $\ \ \ $0.1(25.0) \\ 
$2^3\mbox{S}$ &
1\,152\,842\,738.2(25.2)& 1\,152\,842\,742.87(6)$^d$& $\ \ \ $4.7(25.2) \\
\end{tabular}
\end{table}


\end{document}